\documentstyle[12pt]{article}

\newcommand{\sint}{\sin \theta}        % sin theta
\newcommand{\cost}{\cos \theta}        % cos theta
\newcommand{\nue}{\nu_{\rm e}}         % nu-e
\newcommand{\numyu}{\nu_{\mu}}         % nu-myu
\def\PTP{\em Prog.Theor.Phys.}
\def\PR{\em Phys.Rev.}
\def\PRL{\em Phys.Rev.Letters}
\def\SOKEN{\em Soryushiron Kenkyu} 
\def\KAKU{\em Genshikaku Kenkyu}
\def\PROC{\em Meeting on Beta Decays and Muon Captures 
      in Complex Nuclei}
\def\JETP{\em J.Exptl.Theoret.Phys. (U.S.S.R)}

\begin{document}
%
% -----------------------------------------

\title{PROPOSAL OF NEUTRINO FLAVOR OSCILLATION \thanks{Talk at the Conference on
Neutrino Physics at Ofunato, November 4 - 7, 1998, Ofunato, Iwate, Japan; {\it Proc. of 
the Neutrino Physics Conference at Ofunato}, November 1998, ed. by M. Koga and A. Suzuki, 
Tohoku University, p.1-10}}
\vskip 2em
\author{M. Nakagawa \\
Department of Physics, Meijo University, Nagoya 468-8502, Japan \\
E-mail; mnkgw@meijo-u.ac.jp }
\date{}
\maketitle
\begin{abstract}
A brief survey of the history of the neutrino oscillation is presented. Here emphasis is laid 
on the topics around the first proposal of neutrino flavor oscillation by Nagoya group in 60's. 
Main contents are presented in my opening address " Birth of Neutrino Oscillation" at the 
NOW` 98, Amsterdam, hep-ph/9811358,  but added some detail of the other topics, especially a first 
calculation of decay processes $\mu \to e + \gamma$ and $\nu_2 \to \nu_{1} + \gamma$ in terms of 
neutrino flavor mixing, and some arguments relating to the massive neutrino.
\end{abstract}

{\bf CONTENTS} \\

1. Works on the neutrino mass and oscillation of Nagoya group  \\

2. Proposal of Neutrino Oscillation of a $\nu \leftrightarrow \overline{\nu}$ Transition  \\
\hspace*{1.5cm} Pontecorvo (1957), (1958)  \\

3. Proposal of Flavor Mixing and Flavor Oscillation of Neutrinos \\
\hspace*{1.5cm} Maki, Nakagawa and Sakata (1962) \\
\hspace*{1.5cm} Nakagawa, Okonogi, Sakata and Toyoda (1963) \\
\hspace*{1.5cm} Nakagawa and Toyoda (1964) \\

\hspace*{0.2cm} 3.1 Upper bound on the neutrino mass from the high energy neutrinos \\
\hspace*{1.0cm} 3.2 Beta decays emitting the mixed massive neutrinos  \\
\hspace*{1.0cm} 3.3 Decay processes $\mu \to e + \gamma$ and $\nu_2 \to \nu_{1}+ \gamma$  \\
\hspace*{1.0cm} 3.4 Massive neutrino and inverse beta process  \\

4. Flavor Oscillation of Majorana Neutrino   \\
\hspace*{1.5cm} Pontecorvo (1967) \\
\hspace*{1.5cm} Gribov and Pontecorvo (1969)  \\

5. Summary

\section{\bf Works on the neutrino mass and oscillation of Nagoya group } 

\hspace*{10pt} Works of Nagoya group on the neutrino mass and oscillation in 60's will be given in the 
references, here only the reviews on the studies of neutrino by the group are given. \\

{\bf Reviews on the studies of neutrino by Nagoya group:} \\

\noindent 1) Neutrinos and Sakata - A Personal View -  \\
\hspace*{14pt}M.Nakagawa, {\em Proc. of Neutrino Mass Miniconference}, Telemark, Wisconsin, \\
\hspace*{14pt}Oct. 2-4, 1980. Ed. by V.Barger and D.Cline, p.1-4. \\

\noindent 2) A First Analysis of High Energy Neutrinos in Terms of Neutrino Oscillation  \\
\hspace*{14pt}M.Nakagawa, talk at The International Workshop on Nuclear Emulsion  \\
\hspace*{14pt}Techniques, Nagoya, Japan, June. 12-14, 1998, to be published in Proc.. \\

\noindent 3) Birth of Neutrino Oscillation  \\
\hspace*{14pt}M.Nakagawa, Opening address at the Europhysics NEUTRINO OSCILLATION \\
\hspace*{14pt}WORKSHOP (NOW' 98), 7-9 Sept, 1998, Amsterdam;   \\ 
\hspace*{14pt}Meijo Preprint, 24 Sept. 1998. hep-ph/9811358. \\ 

\section {Proposal of Neutrino Oscillation of a $\nu \leftrightarrow \overline{\nu}$ Transition (1957)}
\begin{flushright}
 B.Pontecorvo (1957), (1958)~\cite{P}
\end{flushright}

* "If the two-component neutrino theory should turn out to be incorrect and if the conservation law of 
neutrino charge would not apply, then in principle neutrino $\to$ antineutrino transitions could take 
place in vacuo"(1957)~\cite{P} just on the analogy to the ${\rm K}^{0}$ to $\overline{\rm K}^{0}$ transition of 
Gell-Mann and Pais~\cite{GP}.  \\

\bigskip

* Define the mixed particles as(1958)~\cite{P} 
\begin{eqnarray}
 \nu &=& \frac{1}{\sqrt{2}}(\nu_{1} + \nu_{2}),              \\
 \overline{\nu} &=& \frac{1}{\sqrt{2}}(\nu_{1} - \nu_{2}),   
\end{eqnarray}
where $\nu_{1}$ and $\nu_{2}$ are, he called, \it truely neutral Majorana particles \em which are mass 
eigenstates.  \\

\bigskip

* A stream of neutral leptons consisting mainly of antineutrinos: 
\begin{equation}
 \overline{\nu} \to \overline{\nu}(50\%) + \nu (50\%),
\end{equation}
will cause a decrease of the capture cross-section of the antineutrinos to the half of the simple 
$\beta$ interaction.  And it was first pointed out the possibility of the observation of the oscillation 
effect on an astronomical scale, which has long been a key concept to solve the solar neutrino problem.

\section{Proposal of Flavor Mixing and Flavor Oscillation of Neutrinos (1962)} 
\begin{flushright}
 Z.Maki, M.Nakagawa and S.Sakata (1962)~\cite{NAGOYA}    \\
\end{flushright}

*{\bf The first proposal of the concept of flavor mixing and flavor oscillation.} \\ 

* The weak interactions of the Sakata model(1955)~\cite{S2}  with a current(Okun'(1958)~\cite{OKN}), 
\begin{equation}
 J_\lambda = (\overline{e}\nu)_\lambda + (\overline{\mu}{\nu})_\lambda + (\overline{n}p)_\lambda + \epsilon(\overline{\Lambda}p)_\lambda,  \label{Okun}
\end{equation} 
have a lepton-baryon symmetry when $\epsilon = 1$;
\begin{eqnarray}
 \nu  \; \; \; &\longleftrightarrow& \;\; p \nonumber  \\ 
 e^{-}         \; \; &\longleftrightarrow & \;\; n     \\                    
 \mu^{-}       \; \; &\longleftrightarrow & \;\; \Lambda ,     \nonumber                  
\end{eqnarray}

which was pointed out by Gamba, Marshak and Okubo (1959)~\cite{GMO}.  The Sakata fundamental baryons 
were assumed as composite particles of leptons and a charged boson responsible for the strong 
interaction~\cite{MNOS}.  \\

* After the confirmation of the two-neutrino hypothesis~\cite{BRKHN}, A new model (Maki, Nakagawa 
and Sakata (1962))~\cite{NAGOYA} was proposed :  \\
(1) Neutrinos should be of 4-component spinors in order to be seeds of the massive baryons.  Consequently, the 
neutrinos $\nu_{1}$ and $\nu_{2}$ to be bound in the baryons should have naturally their own masses.  We called 
these neutrinos as {\bf true neutrinos}\/.   \\
(2) $\nue$ and $\numyu$ coupled to e and $\mu$ in the weak current should be mixing states of $\nu_{1}$ and 
$\nu_{2}$.  We called the neutrinos $\nue$ and $\numyu$ as {\bf weak neutrinos }\/.  
\begin{eqnarray}
 \nue   &=& \cost \; \nu_{1} - \sint \; \nu_{2},  \nonumber  \\ 
 \numyu &=& \sint \; \nu_{1} + \cost \; \nu_{2}.  \label{mixing}            
\end{eqnarray}
The correspondence are as follows: 
\begin{eqnarray}
 \nu_{1} =  \cost \; \nu_{e} + \sint \; \nu_{\mu} \; &\longleftrightarrow& \;\; p \nonumber  \\ 
 \nu_{2} = -\sint \; \nu_{e} + \cost \; \nu_{\mu} \; &\longleftrightarrow & \; X    \\
 e^{-}         \; \; &\longleftrightarrow & \;\; n    \nonumber  \\                    
 \mu^{-}       \; \; &\longleftrightarrow & \;\; \Lambda ,      \nonumber                  
\end{eqnarray}
where we expressed the angle $\theta$ as $\delta$ in the paper.

 The leptonic charged weak current is written as 
\begin{equation}
 j_\lambda = \cost(\overline{e}\nu_{1})_\lambda + \sint(\overline{\mu}{\nu}_{1})_\lambda - 
\sint(\overline{e}\nu_{2})_\lambda + \cost(\overline{\mu}\nu_{2})_\lambda,  \label{lepton}
\end{equation} 
and the baryonic charged weak current is obtained as 
\begin{equation}
 J_\lambda = \cost(\overline{n}p)_\lambda + \sint(\overline{\Lambda}p)_\lambda - 
\sint(\overline{n}X)_\lambda + \cost(\overline{\Lambda}X)_\lambda,    \label{baryon}
\end{equation}   \\
which reproduced the modified current of eq.(\ref{Okun}) suggested by Gell-Mann and L\'{e}vy~\cite{GL}
as
\begin{equation}
  \frac{1}{\sqrt{1+\epsilon^{2}}}(\overline{n}p)_\lambda + 
    \frac{\epsilon}{\sqrt{1+\epsilon^{2}}}(\overline{\Lambda}p)_\lambda.
\end{equation}

\vspace*{1cm}

{  A few remarks: \\

(1) Sakata fundamental baryons are now taken to be quarks.   \\

(2) The structure of the baryonic weak charged current including mixing angle $\theta$ is identical with 
the current involving the Cabbibo angle that is transfered from the mixing angle of neutrinos
%%%%%%%%% Footnote {1} %%%%%%%%%%%%%%%
\footnote[1]{The correspondence was also proposed by Y.Katayama, K.Matumoto, S.Tanaka and E.Yamada, 
{\PTP} \underline{28}, 675 (1962) from a different point of view on neutrinos.}
%%%%%%%%% End of footnote [1} %%%%%%%%%%
.  The proposal 
of the Cabibbo angle was made in 1963~\cite{CAB}. \\

(3) As regards the origin of the mixing angle, it will be still one of the largest problems beyond the standard model.  \\

(4) The fourth baryon X came also naturally into the above correspondence.  But this particle was considered 
as having no seat in the weak current from unknown reason or as being a very large mass particle not yet 
discovered.   Later on, this particle became a candidate for the fourth quark, and was discovered as the 
charm~\cite{CHARM}. } \\

\subsection{Upper bound on the neutrino mass from the high energy neutrinos} \
\begin{flushright}
 Z.Maki, M.Nakagawa and S.Sakata (1962)~\cite{NAGOYA}    \\
\end{flushright}

* The weak neutrinos $\nue$ and $\numyu$ are {\bf not stable}\/ due to the the transmutation $\nue \leftrightarrow \numyu$.
A chain of reactions 

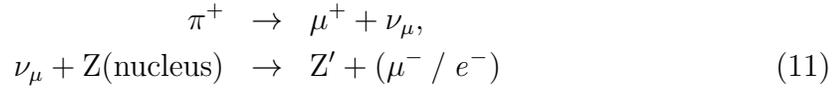
\begin{eqnarray}
  \pi^{+} &\to& \mu^{+} + \numyu ,     \nonumber \\
  \numyu + {\rm Z (nucleus)}\/ &\to& {\rm Z'}\/  + (\mu^{-} {\em \: / \: e^{-}})
\end{eqnarray}
 will take place as a consequence of oscillation.  \\

\bigskip
  
* The (half) oscillation time we have used there was 
\begin{eqnarray}
 T &=& \frac{\pi}{\vert E_{1} - E_{2}\vert}   \nonumber   \\
   &\simeq& 2\pi \frac{pc}{m_{2}c^2}\cdot \frac{M_p}{m_{2}}\cdot 0.7\times 10^{-24}{\rm sec} \:,  
\end{eqnarray}
where assumed as $m_{1} = 0$. (See Appendix A.) \\

\bigskip

* Neutrinos in the two-neutrino experiment by Danby et al.(1962)~\cite{BRKHN}.  \\
Geometry of the neutrino path was taken as 100m, the flight time is 
\begin{equation}
 t_G = \frac{1}{3}\times 10^{-6}{\rm sec}. 
\end{equation}
 Assume for the neutrino beam as
\begin{eqnarray}
 pc &=& 1\:{\rm BeV},      \nonumber  \\
 m_{1}c^{2} &=& 0,                \\
 m_{2}c^{2} &=& x\:{\rm MeV}.  \nonumber
\end{eqnarray}
 Then no observation of $\nu_e$ would mean $ T \ge t_G $, which gives an upper bound 
\begin{equation}
m_{2}c^{2} \leq 3 \cdot 10^{-6}\: {\rm MeV} \; \approx \;{\rm an \; order \; of} \; 1 \; {\rm eV}.
\end{equation} 

\subsection{Beta decays emitting the mixed massive neutrinos  } \
\begin{flushright}
  M.Nakagawa, H.Okonogi, S.Sakata and A.Toyoda (1963)~\cite{NAGOYA}
\end{flushright}

*An emission of a massive neutrino together with a massless neutrino would cause an apparent change in 
the magnitudes of the effective $\beta$ coupling constants and an anomalous kink in the Kurie-plots. 
The $\beta$ interaction is now given as 
\begin{equation}
  -L_{\beta} = \frac{G_{\rm F}}{\sqrt{2}}(\overline{p}n)_{\lambda}\{\cos^{2}\theta (\overline{e}\nu_{1})_\lambda 
 - \cos\theta\sin\theta(\overline{e}\nu_{2})_\lambda \} + {\rm h.c}.
\end{equation}
When the Q-value is so small, the $\beta$ decay emits only the $\nu_{1}$ (in this analysis, we assumed 
$m_{1} \simeq 0 {\rm MeV}\/$ and $m_{2} \simeq 1 {\rm MeV}\/$), whereas the Q-value is large, the decay emits both 
of neutrinos.  \\

   We have used the following formula for the Kurie-plot analysis as
\begin{eqnarray*}
 \sqrt{\frac{N(E)}{C_{F(GT)}pE}} = \left\{(E_{0}-E)^{2} + \epsilon\lbrack (E_{0}-E)^{2} - 
m_{2}^{2}\rbrack^{\frac{1}{2}}(E_{0}-E)\right\}^{\frac{1}{2}},    
\end{eqnarray*}
where
\begin{eqnarray*}
 \epsilon &=& 0 \quad \quad \quad {\rm for\; the\; emisson\; of\; only\; \nu_{1}}\/,      \\
          &=& \frac{\sin^{2}\theta}{\cos^{2}\theta} \quad {\rm for\; the\; emission\; of\; \nu_{1}\; and \; 
              \nu_{2}}\/.      
\end{eqnarray*}                      \\

*  Reported an anomalous kink in the Kurie-plots by Langer group (we called this effect as "Langer effect") 
and also an increase of the magnitudes of the coupling constants with increase of the Q-values~\cite{BETA}. 
 These suggested: \\
\begin{eqnarray}
 m_{2}c^{2} &\simeq& 1\:{\rm MeV},    \nonumber  \\
 \sin {\theta} &\simeq& 0.16 \sim 0.25.
\end{eqnarray}

\bigskip

* The ratio of $N_e$ to $N_{\mu}$ to be observed in the two-neutrino experiment initiating from 
$\pi \to \mu +\nu_{\mu}$ is given in terms of only the mixing angle $\theta$ as
\begin{eqnarray}
 \frac{N_e}{N_{\mu}} &=& \frac{2\sin^{2}\theta \cos^{2}\theta}{\cos^{4}\theta + \sin^{4}\theta}  \nonumber  \\
 &\simeq& \frac{1}{20} \sim \frac{1}{8} \: .
\end{eqnarray}

\subsection{Decay processes $\mu \to e + \gamma$ and $\nu_2 \to \nu_{1}+ \gamma$} 
\begin{flushright}
  M.Nakagawa, H.Okonogi, S.Sakata and A.Toyoda (1963)~\cite{NAGOYA}
\end{flushright}

 Computed the decay processes $\mu \to e + \gamma$ and $\nu_2 \to \nu_{1}+ \gamma$ with diagrams 
involving the weak boson:

\bigskip

* These diagrams can be given in terms of mass (squared) differences of virtual leptons on account of 
cancellations of divergent terms due to the rotation caused by the mixing angle (later on, this was called as 
G.I.M. mechanism~\cite{GIM}).  \\

 $(m_{1}^{2} - m_{2}^{2})/M_{W}^{2}$ \quad for $\mu \to e + \gamma$, \\

 $(m_{\mu}^{2} - m_{e}^{2})/M_{W}^{2}$ \quad for $\nu_2 \to \nu_{1}+ \gamma$.

\bigskip

* The numerical results were
\begin{eqnarray}
 Br(\mu \to e + \gamma ) &\simeq&  10^{-17},      \nonumber   \\
 \tau(\nu_2 \to \nu_{1}+ \gamma) &\simeq&  10^{10}\:{\rm sec},
\end{eqnarray}
under the same parameters $m_{1} = 0, m_{2} = 1 {\rm MeV}, \sin {\theta} \simeq 0.16 \sim 0.25, 
 M_{\rm W} = 1{\rm BeV}$ . 

\bigskip

* For the Feynman integral factor of the $\mu \to e + \gamma$ process, we followed the work of M.E.Ebel 
and F.J.Ernst, {\em Nuovo Cimento} \underline {15}, 173 (1960) by noting that $m_{2}$ plays their cutoff under 
$m_{1}=0$. And for the $\nu_2 \to \nu_{1}+ \gamma$, we calculated it keeping only the leading term. 
(See Appendix B) \\

Diagrams: a gamma is emitted from the following diagrams.

\subsection{Massive neutrino and inverse beta process} 
\begin{flushright}
  M.Nakagawa and A.Toyoda (1964)~\cite{NAGOYA}
\end{flushright}

{\bf A critique}:  \\
  "The idea of massive neutrino is already wrong from the theorem (Lee and Yang) that the cross section of the inverse 
$\beta$ process of a 4-component neutrino will be one half smaller than the one of 2-component neutrino."  \\

 Today this kind of question may be a missunderstanding of the theorem and quite trivial, but it was not necessary. 
Outline of the answer was as follows: The cross section reads in general 
\begin{equation}
 \sigma = \frac{1}{v}\frac{2\pi}{\hbar}\Bigl\{P_{(+)}\vert\langle(+)\vert H_{\rm int}\vert f\rangle \vert^2 
       + P_{(-)}\vert\langle(-)\vert H_{\rm int}\vert f\rangle \vert^2 \Bigr\}({\rm phase \; volume}),
\end{equation}
where $P_{(\pm)}$ are the statistical weights of helicity states $(\pm)$ for the incident neutrino.
For the 2-comp.(massless) neutrino, $P_{(+)} = 0$, $P_{(-)} = 1$ (further $\langle(+)\vert H_{\rm int}
\vert f\rangle = 0$), thus 
\begin{equation}
 \sigma(2\_{\rm comp}) = \frac{2\pi}{\hbar}\Bigl\vert\langle(-)\vert H_{\rm int}\vert f\rangle \vert^2 
({\rm phase \; volume}).
\end{equation}
For the 4-comp.neutrino, to get the 1 : 2 ratio, the following assumptions are needed; \\
(a) $H_{\rm int}$ contains only a left-handed component of neutrino with $m_{\nu} = 0$, \\
(b) the incident neutrino consists of both helicity states with equal weights, which means  $P_{(+)} = 
P_{(-)} = \frac{1}{2}$ and  $\langle(+)\vert H_{\rm int}\vert f\rangle = 0$.  \\
Thus it obtains
\begin{equation}
 \sigma(4\_{\rm comp; unpol}) = \frac{1}{2}\frac{2\pi}{\hbar}\Bigl\vert\langle(-)\vert H_{\rm int}
\vert f\rangle \vert^2 ({\rm phase \; volume}) = \frac{1}{2}\sigma(2\_{\rm comp}). \\
\end{equation}

  The defect of the above conclusion is obviously in the assumption (b).  If the incident neutrinos were produced  
by the interactions with the same nature, the neutrinos with small masses would be predominantly of one helicity 
state as   
\begin{equation}
 P_{(\pm)} = \frac{1}{2}\left(1{\mp}\frac{p_{\nu}}{E_{\nu}}\right),
\end{equation}
and it was shown e.g. for the $\nu + {\rm n} \; \to \; \ell^{-} + {\rm p}$ process
\begin{equation}
 \sigma(\nu+{\rm n}\; \to \;\ell^{-}+{\rm p})=\frac{1}{2}\left(\frac{E_\nu}{p_\nu}+\frac{p_\nu}{E_\nu}
 \right)\sigma_{0}(m_\nu),
\end{equation}
where $\sigma_{0}(m_\nu)$ means a part which becomes equal to $\sigma(2\_{\rm comp})$ at $m_{\nu}=0$ as
\begin{equation}
 \sigma_{0}(m_{\nu}=0) \; = \; \sigma(2\_{\rm comp}).
\end{equation}
 
\section {Flavor Oscillation of Majorana Neutrino (1967)} \
\begin{flushright}
 B.Pontecorvo (1967)~\cite{PG}   \\
 V.Gribov and B.Pontecorvo (1969)~\cite{PG}
\end{flushright}

* In 1967 the violation of the leptonic charge conservation together with 
the violation of the muon charge was proposed by Pontecorvo~\cite{PG}:  \\
\begin{eqnarray}
 \nue \leftrightarrow \overline{\nue},\quad \numyu \leftrightarrow \overline{\numyu}, \quad
 \nue \leftrightarrow \overline{\numyu}, \quad  \overline{\nue} \leftrightarrow \numyu.
\end{eqnarray}

Transition el-neutrino $\leftrightarrow$ mu-neutrino, physically as the flavor oscillation takes place. \\

* The above concept has been formulated by Gribov and Pontecorvo in 1969~\cite{PG}. 
The Majorana mass terms with the massless two-component Dirac neutrinos $\nu_{\rm eL}$ and $\nu_{\mu\rm L}$. 
\begin{equation}
 L_{\rm int} = m_{\rm e\overline{e}}\overline{(\nu_{\rm eL})^{c}}\nu_{\rm eL}
             + m_{\mu \overline{\mu}}\overline{(\nu_{\mu\rm L})^{c}}\nu_{\mu\rm L}
             + m_{\rm e\overline{\mu}}\overline{(\nu_{\rm eL})^{c}}\nu_{\mu\rm L} + {\rm h.c.}.
\end{equation}

* Diagonalization of this Lagrangean leads to the Majorana particles $\phi_{1}$ and $\phi_{2}$ which 
have each eigenmass. The original weak left-handed neutrinos are expressed as
\begin{eqnarray}
 \nu_{\rm eL} = \frac{1}{2}(1+\gamma_{5})(\phi_{1}\cos\xi + \phi_{2}\sin\xi),   \\
 \nu_{\mu\rm L} = \frac{1}{2}(1+\gamma_{5})(\phi_{1}\sin\xi - \phi_{2}\cos\xi),
\end{eqnarray}
where the mixing angle $\xi$ is determined in terms of $m_{\rm e\overline{\mu}}$, $m_{\rm e\overline{e}}$ 
and $m_{\mu\overline{\mu}}$. 

\section {Summary} \

* The motivation to the concept of the neutrino oscillation seems, to me, to consist in {\bf two main streams}. 
{\bf One of them} may be of a strong question on the conservation laws concerning leptonic charges either or both 
of the lepton number and the muon charge that would be violated just in analogy with the established evidence of 
${\rm K}^{0}$ to $\overline{\rm K}^{0}$ transition.  {\bf The other} is in the attempt at model building for a 
unified understanding of the leptons and the hadrons.  A unification of four leptons and the fundamental baryons 
at that time suggested the mixing scheme for neutrinos that would explain successfully the structure of the 
baryonic weak current; the universality of the weak interaction and the smallness of strangeness changing interaction.  
This approach leads to the motivation of the proposal of the concepts of flavor-mixing and -oscillation.  \\

* The stage of the today's neutrino study seems (to me) to be {\bf the third burst} of papers since the first 
proposal of the neutrino flavor oscillation ({\bf 1962}):  \\

{\bf The first one} was in {\bf 1977}, which was fired by the experiment of the SIN $\mu \to e + \gamma$ decay: 
\{W.Dey et al., R-71-06, ETHZ-SIN-ZUERICH, {\em SIN Physics Report} No.1, 9 (Dec 1976)\}. It was the after-gauge 
era where many gauge theoretic calculations on the muon number violation appeared.  \\

{\bf The second one} was around and after {\bf 1980}, fired by the Reines group experiment suggesting a neutrino 
oscillation - it was in no sense definitive: \{E.Pasierb et al., {\PRL} \underline{43}, 96 (1979)\}. Both theoretic 
and experimental works appeared to clarify the oscillation phenomena from low and high energy neutrinos. \\

{\bf The third one} was around and after {\bf 1990}, fired by the dicovery of the so-called MSW mechanism (1986). \\

 May be {\bf the fourth one} today fired by the SUPER-KAMIOKANDE experiment of the atmospheric neutrinos that may 
be the final stage for the determination of the details of neutrino oscillation, and at the same time 
a new first stage for the break-through beyond the standard model of the elementary particles.   \\

%%%%%%%%%%%% Appendix [A] %%%%%%%%%%%%%%%%%
{\it \bf Appendix A: Formulation of Neutrino Oscillation}    \\
 
   Here I present the now well-known formula of the oscillation that we made at that time.  The 
calculation followed to the ${\rm K}_{1}$ and ${\rm K}_{2}$ scheme of Gell-Mann and Pais(1955)~\cite{GP} but 
involving an arbitrary mixing angle as follows. \\

\noindent {\it Time development of ${\nu}_{\mu}$ from pion decay} \rm: 
\begin{eqnarray*}
 \vert {\nu}_{\mu},\:t \rangle = \Bigl\{e^{-iE_{1}t}\sin^{2}{\theta} + e^{-iE_{2}t}\cos^{2}{\theta}\Big\}\vert {\nu}_{\mu}\rangle + \frac{1}{2}\Bigl\{e^{-iE_{1}t} - e^{-iE_{2}t}\Big\}\sin2\theta \vert {\nu}_{e}\rangle.
\end{eqnarray*}

 Detection probability of ${\nu}_{e}$ at t : 
\begin{eqnarray*}
\vert \langle{\nu}_{e}\mid {\nu}_{\mu},\:t \rangle \vert^{2} = \frac{1}{2}\sin^{2}2\theta\{ 1 - \cos(E_{1} 
- E_{2})t\} .
\end{eqnarray*}

 Detection probability of ${\nu}_{\mu}$ at t : 
\begin{eqnarray*}
\vert \langle{\nu}_{\mu}\mid {\nu}_{\mu},\:t \rangle \vert^{2} = \sin^{4}\theta + \cos^{4}\theta + 
\frac{1}{2}\sin^{2}2\theta \cos(E_{1} - E_{2})t .  
\end{eqnarray*}   
\noindent {\it A half oscillation time of the detection probability of ${\nu}_{e}$ }\rm: 
\begin{eqnarray*}
 T = \frac{\pi}{\vert E_{1} - E_{2}\vert}.  
\end{eqnarray*}

 At relativistic limit, under an assumption $m_{2}\neq 0$, and $m_{1}\simeq 0$  \/,
\begin{eqnarray*}
 \vert E_{1} - E_{2}\vert &=& \vert \sqrt{{\bf p}^{2}+m_{1}^{2}} - \sqrt{{\bf p}^{2}+m_{2}^{2}}\; \vert \nonumber  \\
 &\simeq&  p( 1+ \frac{m_{2}^2}{2\, p^2} ) - p          \\
 &=&  \frac{m_{2}^2}{2\,p} \:.                   
\end{eqnarray*}
\hspace*{10pt} Thus
\begin{eqnarray*}
 T &\simeq& \frac{2\pi p}{m_{2}^2}      \\ 
   & = & 2\pi \frac{pc}{m_{2}c^2}\cdot \frac{M_p}{m_{2}}\cdot 0.7\times 10^{-24}{\rm sec} \:,  
\end{eqnarray*}
\hspace*{12pt}where $M_p$ means the proton mass.    \\
\hspace*{12pt}See also, for example, A.K.Mann and H.Primakoff,\PR \em D\underline{15}, 655 (1977);   \\
\hspace*{12pt}S.M.Bilenky and B.Pontecorvo, {\it Physics Report}, \underline{41}, 225 (1978).  \\

%\pagebreak 

%%%%%%%%%%%% Appendix [B] %%%%%%%%%%%%%%%%%
{\it \bf Appendix B: Decay probabilities of $\mu \to e + \gamma$ and $\nu_2 \to \nu_{1}+ \gamma$}    \\
 
  For the $\mu \to e + \gamma$ decay, the relevant part in the matrix element is 
\begin{eqnarray*}
 && \sint\cost\gamma_{\lambda}(1+\gamma_5)\Bigl\{\frac{i{\bf q}- m_1}{q^2 + m_{1}^2} - 
\frac{i{\bf q}- m_2}{q^2 + m_{2}^2}\Bigr\} \gamma_{\rho}(1+\gamma_5) \\
 & = & \sint\cost\gamma_{\lambda}(1+\gamma_5)i{\bf q}\Bigl\{\frac{m_{2}^2 - m_{1}^2}{(q^2 + m_{1}^2)
       (q^2 + m_{2}^2)}\Bigr\}\gamma_{\rho} (1+\gamma_5),
\end{eqnarray*}
where $\bf q$ means an inner product of q and gamma matrices. Then we noted that when $m_1 = 0$ it gives 
the matrix element of M.E.Ebel and F.J.Ernst, {\em Nuovo Cimento} \underline {15}, 173 (1960) 
where {\bf $m_2$ plays just a cutoff} in their calculation. The result was 
\begin{eqnarray*}
 \rho &=& \frac{\Gamma(\mu^{-} \to e^{-} + \gamma)}{\Gamma (\mu^{-} \to e^{-} + \numyu + \overline \nue)} \\ 
   & = & \sin^2\theta\cos^2\theta \frac{27\alpha}{32\pi}\Bigl(\frac{m_{2}^2}{M_{B}^2}\Bigr)^2 \:,  
\end{eqnarray*}
It was after 14 years that the calculation was carried out in the framework of gauge theory that yielded 
\begin{eqnarray*}
 \rho &=& \sin^2\theta\cos^2\theta \frac{3\alpha}{32\pi}\Bigl(\frac{m_{2}^2}{M_{B}^2}\Bigr)^2 
 \quad \quad ({\rm gauge \; theory}) \; ,
\end{eqnarray*}
see for example, T.P.Chen and Ling-Fong Li, {\PRL}, \underline{38}, 381 (1977). Note that the formula we used 
was 9 times larger than the one of gauge theory. \\  
 
 For the $\nu_2 \to \nu_1 + \gamma$ decay processes, we calculated it in the same way as the $\mu \to e + 
\gamma$ decay. As in the case of the $\mu$ decay, the integral of Feynman amplitude is finite where we computed the 
leading part yielding as follows, 
\begin{eqnarray*}
 \Gamma(\nu_2 \to \nu_1 + \gamma) &=& \sin^2\theta \cos^2\theta \frac{\alpha G_{\rm F}^2}{64\pi^4}m_{2}^5
\Bigl(\frac{m_{\mu}^2 - m_{e}^2}{M_{B}^2}\Bigr)^2 \Bigl\{\frac{17}{12} - \ln \frac{M_{B}^2}{m_{\mu}^2}  \\
 && + \frac{m_{e}^2}{m_{\mu}^{2}-m_{e}^2}\ln \frac{m_{\mu}^2}{m_{e}^2}\Bigr\}^2 \:,  
\end{eqnarray*}
where again was assumed $m_1 = 0$.
  The gauge theoretic result is
\begin{eqnarray*}
 \Gamma(\nu_2 \to \nu_1 + \gamma) &=& \sin^2\theta \cos^2\theta \frac{9\alpha G_{\rm F}^2}{2048\pi^4}m_{2}^5
\Bigl(\frac{m_{\mu}^2 - m_{e}^2}{M_{B}^2}\Bigr)^2 \quad \quad ({\rm gauge \; theory}) \:,  
\end{eqnarray*}
see, for example, a general formula given by W.Marciano and A.I.Sanda, {\em Phys. Letters}, \underline{67B}, 
303 (1977). 

%\pagebreak

\end{document}